\documentclass[twocolumn]{aa}
\usepackage{graphicx}
\usepackage{txfonts}
\usepackage{natbib}
\bibpunct{(}{)}{;}{a}{}{,}
\begin{document}
\title{Discovery of the \textit{INTEGRAL} X/$\gamma$-ray transient IGR~J00291+5934: a Comptonised accreting ms pulsar ?}

\titlerunning{Discovery of IGR~J00291+5934}

\author
{S.E. Shaw\inst{1,2} 
\and N. Mowlavi\inst{2,4} 
\and J. Rodriguez\inst{3,2} 
\and P. Ubertini\inst{5}
\and F. Capitanio\inst{5}
\and K. Ebisawa\inst{6,2}
\and D. Eckert\inst{2,4}
\and \\T. J.-L. Courvoisier\inst{2,4} 
\and N. Produit\inst{2,4} 
\and R. Walter\inst{2,4} 
\and M. Falanga\inst{3} 
}
\institute{School of Physics and Astronomy, University of
Southampton, SO17 1BJ, UK 
\and \textit{INTEGRAL} Science Data Centre,
CH-1290 Versoix, Switzerland 
\and CEA Saclay, DSM/DAPNIA/SAp (CNRS FRE 2591), F-91191 Gif Sur
Yvette Cedex, France 
\and Observatoire de Gen\`eve, 51 Chemin des Maillettes,
CH-1290 Sauverny, Switzerland 
\and Istituto di Astrofisica
Spaziale, CNR/INAF, Via Fosso del Cavaliere 100, 00133, Rome, Italy
\and NASA Goddard Space Flight Center, Code 661, Building 2, Greenbelt, MD 
20771, USA}
\offprints{\email{simon.shaw@obs.unige.ch}}

\date{Received now / Accepted then}

\abstract{We report the discovery of a high-energy transient with the IBIS/ISGRI detector on board the \textit{INTEGRAL} observatory. The source, namely \object{IGR J00291+5934}, was first detected on 2nd~December~2004 in the routine monitoring of the IBIS/ISGRI 20--60~keV images.  The observations were conducted during Galactic Plane Scans, which are a key part of the \textit{INTEGRAL} Core Programme observations.  After verifying the basic source behaviour, the discovery was announced on 3rd~December. The transient shows a hard Comptonised spectrum, with peak energy release at about 20~keV and a total luminosity of $\sim 0.9 \times 10^{36}$~erg~s$^{-1}$ in the 5--100~keV range, assuming a distance of 3~kpc.  Following the \textit{INTEGRAL} announcement of the discovery of \object{IGR J00291+5934}, a number of observations were made by other instruments.  We summarise the results of those observations and, together with the \textit{INTEGRAL} data, identifiy IGR~J00291+5934 as the 6th member of a class of accreting X-ray millisecond pulsars.

\keywords{gamma-rays: observations -- pulsars:individual \object{IGR J00291+5934} }
}
\maketitle

\section{Introduction}
\label{sec:intro}
IGR~J00291+5934 was discovered on 2nd~December~2004 \citep{dominique}, during the routine monitoring of IBIS/ISGRI 20--60~keV images of Galactic Plane Scan (GPS) observations at the \textit{INTEGRAL} Science Data Centre (ISDC).  In following GPS observations, on 8th~December, the source flux remained basically stable at $\sim 8 \times 10^{-10}$~erg~cm$^{-2}$~s$^{-1}$ with a marginal monotonic decrease.  However, by 11th~December, the source flux had reduced by around 50\% (see Sec.\ref{sec:disco}).

The day after the discovery the same sky region was observed by the {\it Rossi X-ray Timing Explorer (RXTE)}, which detected a 35~mCrab excess with a coherent pulsation at $\sim$~598.88~Hz (1.67~ms) and pulsed fraction $\sim$~6\%, making IGR~J00291+5934 the fastest known accreting X-ray pulsar \citep{atel353}.  Further analysis of {\it RXTE}/Proportional Counter Array data, showed that the source has an orbital period of $147.412 \pm0.006$~min \citep{atel360}. The {\it RXTE} spectrum was consistent with an absorbed power law with an equivalent absorption column density $N_\mathrm{H}\sim 7\times 10^{21}$~cm$^{-2}$, and a photon index of $\sim 1.7$ \citep{atel353}.   Archival {\it RXTE}/All Sky Monitor data suggested that the source had also entered in outburst in 1998 and 2001, which may indicate that IGR~J00291+5934 has a $\sim 3$ year recurrence time \citep{atel357}.  No such activity was seen in archival \textit{BeppoSAX} and \textit{INTEGRAL} data \citep[e.g.][]{atel362}, although these instruments did not make contemporaneous observations with \textit{RXTE}.  Later observations by the \textit{Chandra} X-ray telescope made a more accurate determination of $N_\mathrm{H}=(2.8 \pm 0.4)\times 10^{21}$~cm$^{-2}$ \citep{atel369}.

Observations at radio and optical wavelengths revealed the presence of a transient counterpart at a position consistent with that of the high-energy source, with possible optical emission features \citep{atel353,atel355,atel356,atel361}.  The most accurate optical position has been reported by \citet{atel354}, ($\alpha$,$\delta$)=($00^{h}29^{m}03^{s}\!.06, +59^{\circ}34\arcmin 19\arcsec\!\!.0)\pm 0\arcsec\!\!.5$; the source is located in the galactic plane, away from the galactic centre at ($l, b$) = (120\fdg0964, -3\fdg1765).

In view of the high-energy behaviour, the presence of pulsations, the short orbital period and other similarities with the object SAX~J1808.4--3658 \citep{1998A&A...331L..25I,1998Natur.394..344W,2001A&A...372..916I} we consider IGR~J00291+5934 to be the 6th member of a class of accreting X-ray binaries with weakly magnetised pulsars.

\subsection{\textit{INTEGRAL}}
\label{sec:obs}

\begin{figure*}[t]
\centering
\includegraphics[width=17cm]{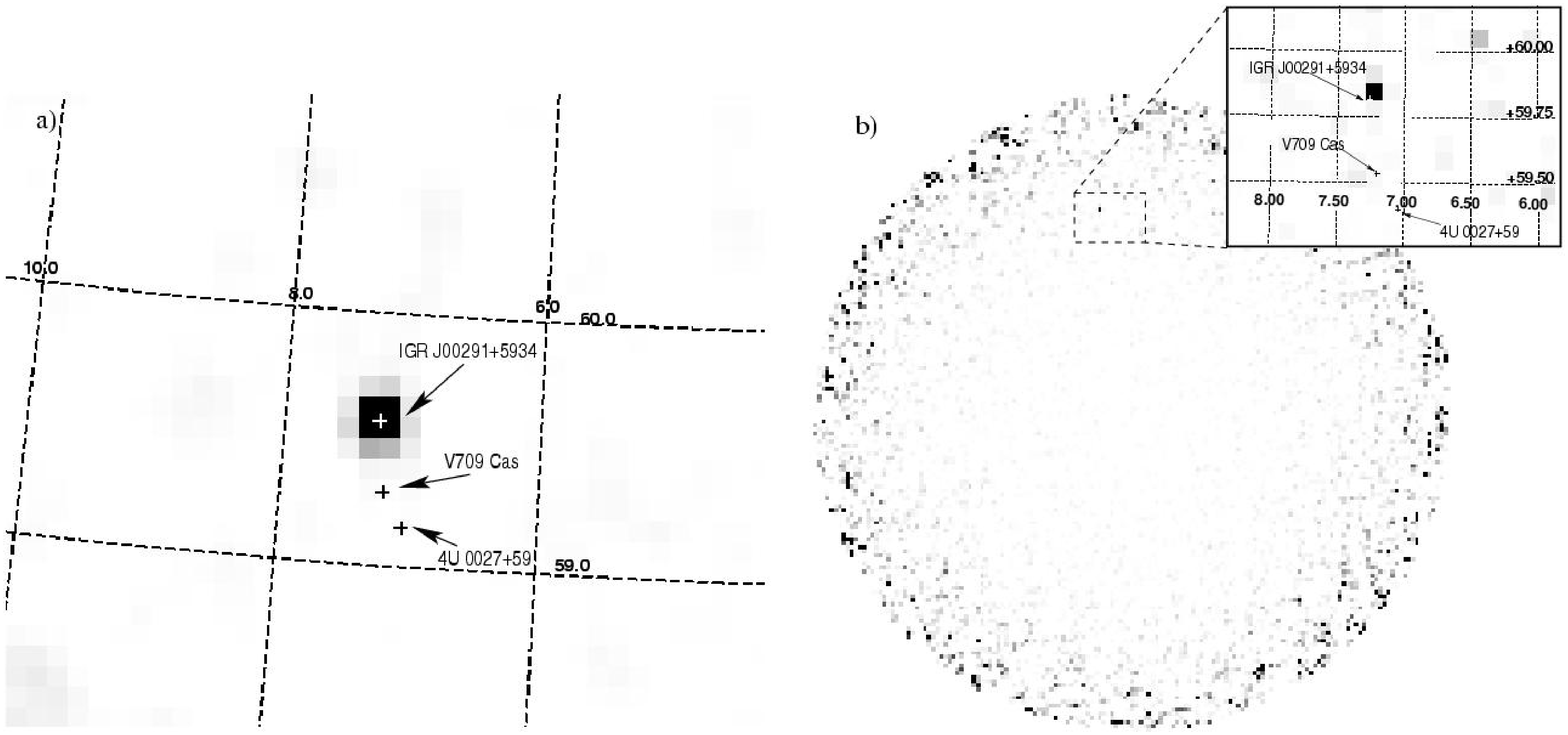}
 \caption{a) Zoomed 15--40~keV ISGRI significance image made from a mosaic of 7 exposures from revolutions 0261 and 0263 (see Table.~\ref{tab:alerts}).   b)  3--10~keV JEM-X1 intensity image, with zoomed inset corresponding to the dashed area (diameter of main image is 10\fdg5).  In both images a squared grey-scale, ranging from 0 to the peak pixel value ($23\sigma$ for ISGRI, 4.4~cps for JEM-X1)  has been used.  It is clear that the \textit{INTEGRAL} detections are co-located with the optical position of IGR~J00291+5934 reported by \citet{atel354} (white cross) and not with the positions of other nearby sources.}
\label{fig:image}
\end{figure*}

The \textit{INTEGRAL} satellite was launched on 17~October~2002 and contains several instruments dedicated to observing the high-energy sky in the 3~keV--10~MeV band \citep{2003A&A...411L...1W}.  The prime instruments for this work are the following: the coded mask imager IBIS/ISGRI, which is sensitive in the 15~keV--1~MeV band and has a large $29^{\circ}\times 29^{\circ}$ field of view \citep{2003A&A...411L.141L,2003A&A...411L.131U}; the X-ray monitor JEM-X, which is sensitive from 3--30~keV and has a 13\fdg2 diameter field of view \citep[although noise towards the rim of the detector limits the usable area for weak sources to the central 10\fdg5; ][] {2003A&A...411L.231L}.  The JEM-X instrument consists of two identical telescopes, but for this work, only the JEM-X1 unit was operational.

Data from the satellite are analysed very quickly after an observation; the ISDC Quick Look Analysis (QLA) pipeline runs continuously on the incoming telemetry \citep{2003A&A...411L..53C}.  Images are produced by QLA in the following energy bands: 3--10~keV and 10--30~keV (JEM-X); 20--60~keV and 60--200~keV (ISGRI).  The images are automatically monitored for new or highly variable sources, which can lead to an automatic alert being issued with a delay of $<$~2~hours from the end of the observation.  All images are also inspected manually by the ISDC Scientist on Duty \citep{munichiqla}.

In this \textit{INTEGRAL} observing period, $\sim$~30\% of the total amount of observing time is split between the Galactic Centre Deep Exposure (GCDE) and the GPS.  The GPS are regular pointings in a saw-tooth pattern along the Galactic Plane, between galactic latitude $b = \pm 10^{\circ}$, conducted every $\sim$~12~days (one \textit{INTEGRAL} revolution is 3~days long).  Each GPS pointing, or {\emph Science Window} (ScW), lasts 2200 seconds and is separated from the next one by 6$^{\circ}$ \citep{2003A&A...411L...1W}. 

\section{Discovery of \object{IGR J00291+5934}}
\label{sec:disco}
The ISDC QLA pipeline first suggested that a new X/$\gamma$-ray source had been discovered in ISGRI images of GPS observations, by an automatic alert issued on 2nd~December~2004 at 09:00:19 UTC \citep{dominique}.  The alert was triggered because a previously unknown source was detected, at a significance $> 10\sigma$, in a 20--60~keV ISGRI image of GPS pointing 0261-2 (ScW 2 of \textit{INTEGRAL} revolution 0261).  This was confirmed at 09:42:03 UTC after the following pointing, 0261-3, by another alert issued at the same sky position.  In both cases the alerts were issued approximately 1 hour and 40 minutes after the end of the pointing (see Table~\ref{tab:alerts} for a summary).

Further QLA images showed that the source persisted in the ISGRI QLA images in 0261-3, and had also been detected in the JEM-X1 instrument in 0261-2 (albeit at a level below that required to trigger an automatic alert).  Due to the progression of the GPS, the source was not in the instrument field of view for the following pointings.  The pointing 0261-2 was also the first pointing of the revolution that could be analysed, since the previous pointing was affected by the passage of the satellite through the Earth's radiation belts.  The next GPS observations, during revolution 0263 on 8th~December~2004, also yielded four detections of the source by ISGRI.  However, in the following GPS pointing, conducted on 11th~December (revolution 0264), the source was observed to have faded in ISGRI, and not detected at all in JEM-X.

After the detection, the source was also observed in an already scheduled  observation of the CasA/Tycho region and was the subject of an \textit{INTEGRAL} ToO observation,  which began on 6th~December.  The analysis of these data is the responsibility of the respective PIs, and will not be discussed here (Falanga, et al., 2005, in preparation).

\begin{table}[th]
  \begin{center}   \caption{Detections of IGR J00291+5934 by the ISDC QLA pipeline, based on ISGRI data from single GPS pointings during \textit{INTEGRAL} revolutions 0261 (2nd~December~2004), 0263 (8th~December~2004) and 0264 (11th~December~2004).  The detected position, angular distance of the source from the spacecraft pointing axis ($\theta$) and 20--60~keV count-rate ($F$) are shown.  The last line shows the detection by JEM-X1 and the 3--10~keV flux.  Pointings marked with * are those where the source was not automatically detected and localised manually.}
\begin{tabular}[h]{lcccc} 
\hline Pointing & UTC Start & ($\alpha$, $\delta$) & $\theta$ & $F_{20-60}$ \\ 
(Rev-ScW) & (hh:mm) &          ($^{\circ}$)          & ($^{\circ}$) &(cps)\\ 
\hline 
0261-2 & 06:43 & (7.26, 59.57) & 3.3  & 7.5$\pm 0.5$\\
0261-3 & 07:23 & (7.24, 59.58) & 6.7  & 6.1$\pm 0.5$ \\
0261-4*& 08:03 & (7.3, 59.57)  & 12.2 & 7.9$\pm 1.2$ \\
0263-1*& 06:31 & (7.3, 59.57)  & 10.2 & 5.5$\pm 0.8$\\
0263-2 & 07:10 & (7.28, 59.58) & 6.7  & 4.3$\pm 0.5$\\
0263-3 & 07:50 & (7.27, 59.58) & 7.6  & 4.6$\pm 0.6$\\
0263-4*& 08:30 & (7.3, 59.57)  & 11.8 & 5.3$\pm 1.0$\\
0264-2*& 06:18 & (7.28, 59.59 )& 10.4 & 1.9$\pm 0.8$\\
0264-3*& 06:58 & (7.4, 59.5)   & 4.6  & 2.2$\pm 0.5$\\
0264-4*& 07:38 & (7.29, 59.55) & 2.3  & 2.6$\pm 0.5$\\
0264-5*& 08:18 & (7.2, 59.5)   & 7.9  & 2.4$\pm 0.7$\\
 \hline
0261-2 & 06:43 & (7.279, 59.568)& 3.3 & 1.6$\pm 0.2$\\
 \hline
\end{tabular} 
\label{tab:alerts} 
\end{center}
\end{table}

\subsection{Analysis of \textit{INTEGRAL} GPS observations}
\label{sec:anal}
The ISGRI GPS pointings, listed in Table~\ref{tab:alerts}, have been analysed using the standard OSA 4.2 software\footnote{The OSA software can be obtained from \emph{www.isdc.unige.ch}.}.  Fig.~\ref{fig:image} shows the location of the source in an ISGRI mosaic image, made from all 7 pointings of revolutions 0261 and 0263, and the single JEM-X detection.  In both instruments the source is very clearly identified with the optical position of \citet{atel354}.  Within the ISGRI mosaic image the HMXB \object{3A 0114+650} is also detected; this gives confidence that some problem with the spacecraft pointing is not responsible for falsely identifying a new source and that IGR~J00291+5934 is not a misidentification of another nearby object.
  
The most significant detection of IGR~J00291+5934 was during the first pointing, 0261-2, and it was possible to construct a composite JEM-X1 and ISGRI spectrum.  However, the sensitivity of ISGRI is such that it is hard to constrain a physical fit to this source on the strength of just one pointing.  A simple absorbed power-law, with the value of $N_{H} = 0.2\times 10^{22}$~cm$^{-2}$ fixed, gives a reduced $\chi^{2}_{\nu} = 1.1$.  The photon index is $1.81\pm0.13$ corresponding to a 5--50~keV flux of $8.3_{-2.6}^{+3.7} \times 10^{-10}$~erg~cm$^{-2}$~s$^{-1}$ in agreement with \citet{atel353}.

The broad band count-rates of IGR~J00291+5934 are also noted in Table~\ref{tab:alerts}.  These show that the flux faded slightly during the course of $\sim$~10~days, as reported by \citet{atel365}.  Note that the uncertainties on flux measurements with ISGRI increases with the off-axis angle, $\theta$, as the source flux is not fully coded by the mask.

\begin{figure}[h]
 \centering
 \resizebox{\hsize}{!}{\includegraphics{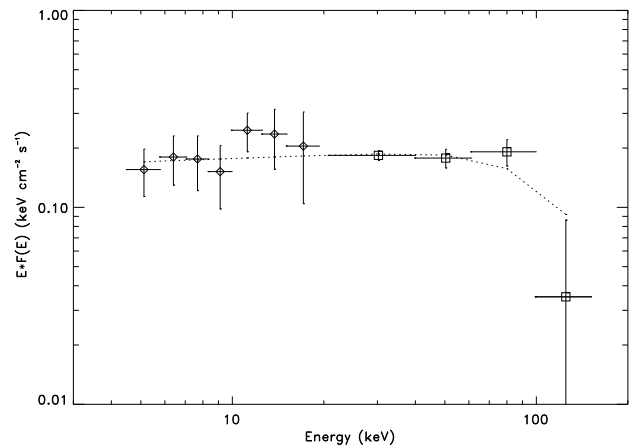}}
 \caption{X/$\gamma$-ray spectrum of IGR~J00291+5934 made with JEM-X1 from observation 0261-2 (circles), and an average of ISGRI observations (squares, see text for data selection).  The line shows a simultaneous fit to the two sets of data with the {\tt CompST} model.}
 \label{fig:spec}
\end{figure}

If IGR~J00291+5934 is an accreting ms pulsar, then a power-law does not necessarily describe the physics of the source and it is interesting to investigate the possibility of a Comptonised spectrum.  To increase the high-energy statistics, an average ISGRI spectrum was made, using the method described in \citet{rodriguez}, from those pointings in Table~\ref{tab:alerts} with $\theta < 10^{\circ}$; this was added to the JEM-X1 spectrum from 0261-2.  A reasonable fit was obtained with the {\tt CompST} model \citep{1980A&A....86..121S}: $\chi^{2}_{\nu} = 0.7$, electron temperature $kT = 25^{+21}_{-7}$~keV and optical depth $\tau = 3.6^{+1.0}_{-1.3}$; the 5--100~keV flux is  $8.5^{+0.5}_{-0.2} \times 10^{-10}$~erg~cm$^{-2}$~s$^{-1}$.  However, it should be noted that a simple power-law also gives a valid fit, $\chi^{2}_{\nu} = 1.0$, albeit with a softer photon index of  $2.05\pm 0.10$.  The drop in the flux above 100~keV confirms the presence of a cut-off in the spectrum at high energies (Fig.~\ref{fig:spec}).

Unfortunately, the source was not bright enough to extend the high-energy spectral information with the \textit{INTEGRAL} spectrometer \citep[SPI;][]{2003A&A...411L..63V}.  The source was, however, detected at a corresponding position to the other instruments, ($\alpha$,$\delta$)=(6\fdg3,+59\fdg8) $\pm$ 1\fdg0, and the measured 20--60~keV flux of $(5.3 \pm 1.6) \times 10^{-10}$~erg~cm$^{-2}$~s$^{-1}$ (assuming a photon index of 1.8) is in good agreement with ISGRI.

The ISGRI data have also been searched for pulsations at the 1.67~ms period of \citet{atel353}.   To maximize the signal to noise ratio, the 20--60~keV band was used, noisy pixels were removed and only those pixels that were fully illuminated by IGR~J00291+5934 were selected.  The event arrival time was corrected to the solar barycentre and a folded analysis was conducted around the known pulse-frequency of 598.88 Hz.  An upper limit for the pulsed amplitude at 598.88 Hz was found to be $\sim$~20\%, which is consistent with the value of ~6\% by \citet{atel353}.

\section{Discussion}
We report the discovery of the fastest known accreting X-ray pulsar, IGR~J00291+5934, with \textit{INTEGRAL}.   It is likely that IGR~J00291+5934 is a low mass X-ray binary (LMXB) system containing a NS pulsar that has been spun up by accretion of material, from a companion star, via an accretion disc.  IGR~J00291+5934 is one of the fastest X-ray pulsars discovered to date and is second only to PSR~B1937+21 \citep[an isolated pulsar showing radio and X-ray pulsations at $P\sim 1.57$~ms; ][]{1982Natur.300..615B,2001ApJ...554..316T}.  The absorbing column measured by \citet{atel369} with \textit{Chandra} is approximately a factor of two lower than the estimate of the galactic  total on the same line of sight \citep[$N_\mathrm{H}\sim 4.5\times 10^{21}$~~cm$^{-2}$;][]{1990ARA&A..28..215D}.  Given that the source is 120$^{\circ}$ from the Galactic Centre and assuming that the galactic disk has a radius of 13~kpc, with the Earth 8.5~kpc from the centre then the average density of absorbing material in the source direction is $\sim$~0.2~cm$^{-3}$.  Assuming no local absorption puts an upper limit on the source distance of $\sim$~3.3~kpc.  Although this is a highly simplistic argument, it seems likely that the source is reasonably local.  It should also be noted that the Perseus arm of the Milky Way, at $l = 120^{\circ}$, is located approximately 2.5~kpc away \citep{1993ApJ...411..674T}.  Using 3~kpc as an estimate of the source distance, and the 5--100~keV flux quoted in Sec.~\ref{sec:anal}, gives a luminosity for the source of $\sim 0.9 \times 10^{36}$~erg~s$^{-1}$.

The other five known ms X-ray pulsars (\object{SAX J1808.4-3658}, \object{XTE J1751-305}, \object{XTE J0929-314}, \object{XTE J1807-294}, \object{XTE J1814-338}), are believed to be old NS (age $\sim 10^{9}$~years), with moderately weak magnetic field, $B\sim 10^{8}$~G \citep[see e.g.][]{2003Natur.424...42C,2003Natur.424...44W}. In fact all of them are transient systems with short orbital periods, accreting at very low rates.    This implies that the magnetic field of the NS is very weak, which is also suggested by the perceived age of the systems \citep{1998ApJ...506L..35H,2002ApJ...576L..49T}.   The \textit{INTEGRAL} high-energy spectral information is limited with this very small data set, but is consistent with properties of other ms X-ray pulsars, with $\tau \sim 3$ and $k_{B}T \sim 20$~keV.  The upper limit on the high-energy pulsations, coupled with the peak energy release at $\sim$~24~keV, could be consistent with a Comptonised flux being emitted from a hot plasma near the inner part of the disc.  Some of this plasma may be channeled towards the NS by the magnetic field, resulting in the pulsed part of the spectrum.  In this model, the seed photons would be supplied by the cold intermediate part of the disc rather than the higher temperature NS black-body emission.

It is remarkable that in all of these objects the pulse fraction is of the same order ($\sim 6$\%). IGR~J00291+5934, shares many common characteristics with the other objects, particularly SAX~J1808.4--3658. The latter  has a relatively similar orbital period of $\sim$~2~hours \citep{1998Natur.394..346C}, a $\sim$~2~year recurrence time of the outburst and is also the only other known ms pulsar for which a radio counterpart was detected during outburst \citep{1999ApJ...522L.117G}.  It is reasonable to assume that \object{IGR J00291+5934} is an old NS that has been spun up by the accretion of material, with a magnetic field of the same order as the other ms pulsars.  The spectral analysis of \object{IGR J00291+5934} reveals that this source is more similar to XTE~J1814--338 than the 3 others for which evidence of black body radiation have been seen. However, this may simply be due  to the high lower boundary of the JEM-X detector. This is reinforced by the fact that black body emission has been detected in SAX~J1808.4--3658 with detectors allowing a broader coverage towards the low energies \citep[e.g.][]{2003Natur.424...44W}.

The spectral analysis of these objects reveal that they are not so different to other NS/pulsar LMXBs, in the sense that the emission processes are thought to originate through thermal + Comptonised processes.  It is therefore quite puzzling that only some of these systems show persistent coherent pulsations at the NS spin period.   The fact that all ms X-ray pulsars have very short orbital periods may be a clue to why these systems do or do not show persistent pulsations \citep[see recent review by ][]{aph0501264}. Despite being the fastest accreting ms pulsar to date, it is interesting to note that the period remains significantly higher than 1~ms.

\begin{acknowledgements}
This paper is based on observations made with the ESA \textit{INTEGRAL} project.  The authors thank A.Paizis for useful comments in the preparation of this paper.  SES thanks PPARC for financial support.  The useful comments and timely response of the anonymous referee were greatly appreciated.
\end{acknowledgements}

\bibliographystyle{aa}
\bibliography{igr00291}

\end{document}